\documentstyle[preprint,epsf,eqsecnum,aps]{revtex}

\begin{document}
\def\appls{\hbox{$<$\kern-.75em\lower 1.00ex\hbox{$\sim$}}}
\draft
\tightenlines
\title{$\sigma(770)$ RESONANCE AND THE BREAKING OF \\
SCALE AND CHIRAL SYMMETRY IN EFFECTIVE QCD}
\author{M. Svec\footnote{electronic address: svec@hep.physics.mcgill.ca}}
\address{Physics Department, Dawson College, Montreal, Quebec, Canada H3Z 1A4\\
and\\
Physics Department, McGill University, Montreal, Quebec, Canada H3A 2T8}
\maketitle
\begin{abstract}
CERN measurements of $\pi^- p \to \pi^-\pi^+ n$ on polarized target at 17.2 GeV/c enable experimental determination of partial wave production amplitudes below 1080 MeV. The measured $S$-wave transversity amplitudes provide evidence for a narrow scalar resonance $\sigma(770)$. The assumption of analyticity of production amplitudes in dipion mass allows to determine $S$-wave helicity amplitudes $S_0$ and $S_1$. The helicity flip amplitude $S_1$ is related to $\pi^- \pi^+ \to \pi^- \pi^+$ scattering. There are four "down" solutions $(1, \overline 1)$, $(2, \overline 1)$, $(1, \overline 2)$ and $(2, \overline 2)$ selected by the unitarity in $\pi^- \pi^+ \to \pi^- \pi^+$ scattering. Ellis-Lanik relation between the mass $m_\sigma$ and partial width $\Gamma( \sigma \to \pi^- \pi^+)$ derived from an effective QCD theory with broken scale and chiral symmetry selects solutions $(1, \overline 1)$ and $(1, \overline 2)$ and imparts the $\sigma(770)$ resonance with a dilaton - gluonium interpretation. Weinberg's mended symmetry selects solutions $(1, \overline 1)$ and $(2, \overline 1)$. The combined solution $(1, \overline 1)$ has $m_\sigma = 769 \pm 13$ MeV and $\Gamma_\sigma = 154 \pm 22$ MeV. The observed $\sigma - \rho^0$ degeneracy leads to new relations between gluon condensate and pion decay constant and $SU(2)$ chiral condensate. The two relations are well satisfied. Ellis-Lanik relation relates the existence of $\sigma(770)$ to breaking of scale symmetry in QCD. Interaction of chirally invariant and narrow $\sigma(770)$ with chirally noninvariant QCD vacuum in $\pi^- \pi^+ \to \pi^- \pi^+$ scattering is proposed as a possible mechanism for the metamorphosis of $\sigma(770)$ into a broad resonant $S$-wave structure in $\pi^- \pi^+ \to \pi^- \pi^+$ related to spontaneous breaking of chiral symmetry in QCD. We comment on possible connections of $\sigma(770)$ to cosmological contstant and QCD/String duality in AdS/CFT brane world model. 
\end{abstract}

\pacs{} 
\section{Introduction}

In 1968, Mack, Wilson and Gell-Mann recognized that scale invariance is a broken symmetry of strong interactions\cite{mack68,wilson69,gellmann69,carruthers71}. In 1969, Salam and Strathedee showed that to formulate a broken chiral as well as scale symmetry within an effective lagrangian approach one has to assume the existence of a chirally invariant scalar field $\sigma$ of dimension 1\cite{salam69}. In 1970, Ellis proposed to identify this scalar field with the $\sigma(700)$ meson\cite{ellis70} the existence of which was suggested by earlier measurements of forward - backward asymmetry in
$\pi^- p \to \pi^- \pi^+ n$\cite{hagopian63,islam64,patil64,durand65,baton65}. The scalar meson dominance of the trace of the energy-momentum tensor (also referred to as a partially conserved dilatation current) has been used to study the couplings of the $\sigma(700)$ 
meson\cite{ellis70,crewther72,chanowitz72,chanowitz73}.

With the advent of QCD in 1970's it has been recognized that the quantization of QCD lagrangian leads to breaking of scale invariance in QCD. The anomalous breaking of scale symmetry results in QCD scale anomaly which was 
shown\cite{adler77,collins77,nielsen77} to have the form

\begin{equation}
\partial_\mu J^\mu_{scale}=\theta^\mu_\mu= {\beta(g) \over {2g}}G_{\mu \nu}G^{\mu \nu} + (1+\gamma_m) \sum_{u,d,s} m_q \overline q q
\end{equation}

\noindent
Here $G_{\alpha \beta}$ and $q$ are the gluon field strength and the quark field with running mass $m_q$. $\beta$ and $\gamma_m$ are the Gell-Mann $\beta$-function and quark anomalous dimension\cite{bertlmann96,yndurain99,pokorski00}. The summation over colour is understood. $\theta^\mu_\mu$ is the trace of the energy - momentum tensor.

In the absence of a technology to solve the fundamental QCD theory and find the hadron spectrum and the interactions of the composite states, we use the effective lagrangian method to study the hadron dynamics at low energies\cite{pokorski00,dobado97}. The basic ingredient in constructing effective lagrangians is anomaly matching. The effective lagrangian must posses not only the symmetries but also the anomalies of the original QCD theory\cite{pokorski00}.

In 1981, Schechter suggested that a scalar gluonium field $\sigma$ plays a major role in effective QCD lagrangian through its connection to the QCD trace 
anomaly\cite{schechter80,salomone81}. Effective lagrangians with such dilaton-gluonium field $\sigma$ were subsequently examined from various aspects in a series of 
studies\cite{gomm86,cornwall84,cornwall85,migdal82,lanik84,ellis85,ellis86}. 

In 1985, Ellis and Lanik\cite{ellis85} constructed an effective QCD lagrangian with broken scale and chiral symmetry in which the dilaton-gluonium scalar field $\sigma (x)$ is related to the scalar gluonic current $H(x)$ by a relation

\begin{equation}
H(x) \equiv - {\beta (g) \over {2g}} G_{\mu \nu}G^{\mu \nu} = 
{m^4_\sigma \over {16 \pi H_0}} \sigma(x) ^4  
\end{equation}

\noindent
In Eq. (1.2) $m_\sigma$ is the $\sigma$ meson mass and $H_0 = <0|H|0>$  is related to gluon condensate $G_0$ 

\begin{equation}
G_0=<0|(\alpha_s / \pi) G_{\mu \nu} G^{\mu \nu} |0>
\end{equation}

\noindent
by an approximate relation\cite{ellis85}

\begin{equation}
H_0 = {9 \over 8} G_0
\end{equation}

\noindent 
The gluon condensate $G_0$ parametrizes the nonperturbative effects of QCD and is related to the energy density of QCD vacuum. The relation (1.2) is unique to Ellis-Lanik lagrangian. Starting with the Salam-Strathedee chirally invariant field $\sigma(x)$, it is the result of  matching of the QCD trace anomaly in gluonic sector with the trace of the energy-momentum tensor of the $\sigma$ field\cite{callan70} and low-energy theorems for scalar gluonic current $H(x)$\cite{novikov81}. From their lagrangian Ellis and Lanik derived the following relations for $\sigma$  decay widths

\begin{equation}
\Gamma(\sigma \to \pi^- \pi^+) = { m^5_\sigma \over{ 48 \pi G_0}}
\end{equation}

\begin{equation}
\Gamma(\sigma \to \gamma \gamma)= {3 \over 4} \Bigl( {R \alpha \over {8 \pi^2}} \Bigr)^2 \Gamma(\sigma \to \pi^- \pi^+)
\end{equation}

\noindent
where $R=\sigma(e^-e^+ \to hadrons)/ \sigma(e^-e^+ \to \mu^- \mu^+)$. 

The appearance of the Gell-Mann function $\beta$ in the scale anomaly (1.1) reflects the QCD confinement. In the Ellis-Lanik lagrangian the $\sigma$ field codes the QCD confinement  which is often a missing feature in other effective QCD lagrangians.

The CERN measurements of $\pi^- p \to \pi^- \pi^+ n$ and $\pi^+ n \to \pi^+ \pi^- p$ on polarized targets reopened the question of existence of the $\sigma(750)$ meson. These measurements allow a model independent determination of normalized production amplitudes, including the two $S$-wave transversity amplitudes. Evidence for a narrow $\sigma(750)$ resonance was found in amplitude analyses of CERN data on $\pi^- p \to \pi^- \pi^+ n$ at 17.2 GeV/c in the mass range 600 - 900 MeV and on $\pi^+ n \to \pi^+ \pi^- p$ at 5.98 and 11.85 GeV/c in the mass range 580 - 980 MeV\cite{svec92,svec96,svec97,svec98}. Further evidence was found recently in amplitude analysis of  measurements $\pi^- p \to \pi^- \pi^+ n$ on polarized target at 1.78 GeV/c at ITEP\cite{alekseev99}.

Our new amplitude analysis\cite{svec02} of the CERN measurements of  $\pi^- p \to \pi^- \pi^+ n$ on polarized targets at 17.2 GeV/c and momentum transfer $-t$ = 0.005 - 0.20 $(GeV/c)^2$ extends the mass range to 580 - 1080 MeV and allows to study the effects of $\sigma(750) - f_0(980)$ interference. There are two solutions for the unnormalized moduli $|\overline S|^2 \Sigma$ and $|S|^2 \Sigma$ of the two $S$-wave transversity amplitudes $\overline S$ and $S$ corresponding to recoil nucleon transversity "up" and "down" relative to the scattering plane. Here $\Sigma = d \sigma / dt dm$ is the integrated cross-section. Both moduli in both solutions exhibit a resonant behaviour around 750 - 780 MeV. 

In our analysis\cite{svec02} we supplement the CERN data with an assumption of analyticity of production amplitudes in dipion mass. Analyticity allows to parametrize the transversity amplitudes $S$ and $\overline S$ as a sum of Breit-Wigner amplitudes for $\sigma(750)$ and $f_0(980)$ with complex coefficients and a complex background. Next we performed simultaneous fits to the moduli $|S|^2 \Sigma$ and $|\overline S|^2 \Sigma$ in the four solution combinations $(1, \overline 1)$, $(2, \overline 1)$, $(1, \overline 2)$ and $(2, \overline 2)$. In each solution combination we obtained two fits, A and B, with the same resonance parameters for $\sigma(750)$ and $f_0(980)$ and the same $\chi^2 /dof$. The average values of $\sigma$ mass and width are $m_\sigma = 778 \pm 16$ MeV and $\Gamma_\sigma = 142 \pm 33$ MeV.

The transversity amplitudes $S$ and $\overline S$ are linear combinations of nucleon helicity nonflip amplitude $S_0$ and nucleon helicity flip amplitude $S_1$ corresponding to $a_1$ and $\pi$ exchange in the $t$-channel, respectively. These amplitudes are physically interesting since the residue of the pion pole in $S_1$ is related to the $S$-wave partial wave in $\pi^- \pi^+ \to \pi^- \pi^+$ scattering. The residue of the $a_1$ pole in $S_0$ is related to the $S$-wave partial wave in $\pi^- a^+_1 \to \pi^- \pi^+$ scattering.  

Analyticity imparts the fitted transversity amplitudes with an absolute phase. This allows to determine the helicity amplitudes from the fitted transversity amplitudes. Due to a sign ambiguity there are two solutions, "up" and "down" in each solution combination and in each fit A and B. In the "down" solution the $\sigma(770)$ is suppressed in the flip amplitude $S_1$ while it dominates the nonflip amplitude $S_0$. In the "up" solution the situation is reversed. In Ref.~\cite{svec02} we show that unitarity in $\pi^- \pi^+ \to \pi^- \pi^+$ scattering excludes the "up" solution. Furthermore, the "down" solution - and thus the evidence for the narrow $\sigma(770)$ - is in agreement with unitarity in both $\pi^- \pi^+ \to \pi^- \pi^+$ and $\pi^- a^+_1 \to \pi^- \pi^+$ scattering.

In the "down" solution the narrow $\sigma(770)$ manifests itself as a broad resonance centered at $\sim$ 720 MeV in the helicity flip mass spectra $|S_1|^2$ in solution combinations $(1, \overline 1)$ and $(2, \overline 1)$. In contrast, the contribution of $\sigma(770)$ is small in $|S_1|^2$ in combinatins $(1, \overline 2)$ and $(2, \overline 2)$.

The dual manifestation of the $\sigma(770)$ as a narrow resonance in nonflip amplitude $S_0$ and as a broad resonance in the flip amplitude $S_1$ is connected to symmetries of QCD. In this paper we show that Ellis - Lanik relation (1.5) selects the solution combinations $(1, \overline 1)$ and $(1, \overline 2)$. Weinberg's mended symmetry of spontaneously broken chiral 
symmetry\cite{weinberg90}  selects solutions $(1, \overline 1)$ and $(2, \overline 1)$. The combined preferred solution is thus $(1, \overline 1)$  for which $m_\sigma = 769 \pm 13$ MeV and $\Gamma_\sigma = 154 \pm 22$ MeV. 

The agreement with Ellis-Lanik relation imparts the $\sigma(770)$ resonance with a dilaton-gluonium interpretation and relates its existence to breaking of scale symmetry in QCD. At the same time, the degeneracy of $\sigma(770)$ mass with $\rho(770)$ and the metamorphosis of the narrow $\sigma(770)$ into a broad $S$-wave structure in $\pi^- \pi^+ \to \pi^- \pi^+$ scattering is in agreement with Weinberg's mended symmetry which predicts $m_\sigma = m_\rho$ with a broad width\cite{weinberg90}. We suggest that the broad resonant $S$-wave structure in $\pi^- \pi^+ \to \pi^- \pi^+$ scattering associated with the spontaneous breaking of chiral symmetry is due to the interaction of the chirally invariant and narrow $\sigma(770)$ with the chirally noninvariant QCD vacuum in $\pi^- \pi^+ \to \pi^- \pi^+$ scattering.

The paper is organized as follows. In Section II. we use the most recent values of gluon condensate to show that Ellis-Lanik relation (1.5) selects the solutions $(1, \overline 1)$ and $(1, \overline 2)$. In Section III. we show how Weinberg's mended symmetry selects the solutions $(1, \overline1)$ and $(2, \overline 1)$ and comment on the role of QCD vacuum in hadron reactions. The observed degeneracy of $\sigma(770)$ and $\rho^0(770)$ is used in Section IV. to derive new relations between gluon condensate and pion decay constant, and between chiral condensate and gluon condensate. We find that these relatons are well satisfied. The paper closes with a summary and remarks in Section V.

\section{Dilaton - gluonium structure of $\sigma (770)$ resonance}

The moduli of the $S$-wave transversity amplitudes $\overline S$ and $S$ measured in $\pi^- p \to \pi^- \pi^+ n$ at 17.2 GeV/c on polarized target are shown in Figure 1. The figure shows the two solutions for $|\overline S|^2 \Sigma$ and $|S|^2 \Sigma$ and the results of simultaneous fits to the four solution combinations $(1, \overline 1)$, $(2, \overline 1)$, $(1, \overline 2)$ and $(2, \overline 2)$. Two fits, A and B, give essentially identical curves, the same resonance parameters for $\sigma(750)$ and $f_0(980)$, and the same $\chi^2 /dof$\cite{svec02}. The transversity "up" amplitude $|\overline S|^2 \Sigma$ clearly resonates below 800 MeV while the transversity "down" amplitude $|S|^2 \Sigma$ shows a broader structure around 800 MeV.

In Table I we present the mass $m_\sigma$ and the width $\Gamma_\sigma$ of the $\sigma$ resonance for the best fits in each solution combination. The measured width $\Gamma_\sigma$ is the total width of the $\sigma$ decays. To compare our results with the Ellis-Lanik relation (1.5) we need partial width $\Gamma(\sigma \to \pi^- \pi^+) = \xi \Gamma_\sigma$ where $\xi$ is the fraction of charged $\pi^- \pi^+$ decays. As we shall see below, the partial width 
$\Gamma( \sigma \to \gamma \gamma)$ is small. We also assume that $4 \pi$ decays of $\sigma$ are suppressed. Then isospin conservation requires that $\xi = {2 \over 3}$. With this value of $\xi$ we then calculate the value $G_0$ of the gluon condensate from Ellis-Lanik relation for each solution combination. 

The numerical values of $G_0$ were estimated originally by the ITEP group\cite{shifman79} to be $G_0 \approx 0.012$ (GeV)$^4$ or up to $G_0 \approx 0.030$ (GeV)$^4$ in later calculations\cite{bell81,bradley81}. More recent estimates are $G_0 = 0.023 \pm 0.003$ (GeV)$^4$ \cite{narison97} and $G_0 = 0.016 \pm 0.010$ (GeV)$^4$\cite{yndurain99b}. The latest estimate\cite{bertolini98,fabbrichesi00,bertolini01} based on analysis of the $CP$ violating ratio $\epsilon ' / \epsilon$ is  $G_0 = 0.0123 \pm 0.0011$ (GeV)$^4$. The most recent estimate based on charmonium sum rules\cite{ioffe02} gives a similar result $G_0 = 0.011 \pm 0.009$ (GeV)$^4$.  This new ITEP value includes a contribution from $< G^3 >$ condensate in dilute instanton model\cite{ioffe02}. In this paper we will use the average value of the two last results

\begin{equation}
G_0 = 0.0117 \pm 0.0050\ {\rm GeV}^4
\end{equation}

The comparison of $G_0$ in (2.1) with the results for $G_0$ from Ellis -Lanik relation in Table I shows the best agreement is for solution combinations $(1, \overline 1)$ and $(1, \overline 2)$. The Ellis-Lanik relation thus selects the solutions $(1, \overline 1)$ and $(1, \overline 2)$ and imparts the $\sigma(770)$ resonance with a dilaton-gluonium interpretation.    
    
We can now use the relation (1.6) to calculate the decay width $\Gamma(\sigma \to \gamma \gamma )$. Assuming $R=5$ and $\xi = {2 \over 3}$ we obtain results shown in Table I. We see that $\Gamma(\sigma \to \gamma \gamma )$ is too small for $\sigma(770)$ to be observed in $\gamma \gamma$ decays. This result is in full agreement with the measurements of reactions 
$\gamma\gamma \to \pi^+\pi^-$\cite{berger84,aihara86,blinov92,behrend92} and 
$\gamma\gamma \to \pi^0\pi^0$\cite{marsiske90,oest90} which found no evidence of resonance structure below $f_0(980)$. The fact that $\sigma(770)$ is not observed in these reactions  strongly suggests a gluonium interpretation of this state since gluons do not couple directly to photons. These experiments support the Ellis-Lanik identification of the scalar dilaton field $\sigma$  with a scalar gluonium which they based solely on the dominance of the scalar gluonic current $H(x)$ by the field $\sigma$\cite{ellis85}.

A theoretical support for the gluonium interpretation of $\sigma(770)$ comes from recent studies of QCD sum rules. It was shown by Elias and collaborators in \cite{elias98} that QCD sum rules require that a scalar state with a mass below 800 MeV has a narrow width. Several analyses using QCD sum rules concluded that the scalar states with a mass below $f_0(980)$ are non-$q\overline{q}$ states\cite{narison98,steele99}. The QCD sum rules are thus consistent with the identification of the narrow $\sigma(770)$ resonance with the lowest mass scalar gluonium.

\section{Weinberg's Mended Symmertry}

In 1990, Weinberg showed that even where a symmetry is spontaneously broken it can still be used to classify hadron states\cite{weinberg90}. Such mended symmetry leads to a quartet of particles with definite mass relations and C-parity. He showed that $\pi$, $\sigma$, $\rho$ and $a_1$ mesons corresponding to spontaneously broken chiral symmetry of strong interactions occupy the quartet. Weinberg's mended symmetry predicts $m_\sigma = m_\rho$ in agreement with our results for $m_\sigma$ shown in Table I.

Central to Weinberg's mended symmetry are current algebra and superconvergence sum rules for helicity amplitudes. In 1968, Gilman and Harari studied saturation of such sum rules in a framework of a linear sigma model\cite{gilman68}. They found that $m_\sigma = m_\rho$ and $\Gamma_\sigma = {9\over{ 2} }\Gamma_\rho$ which gives $\Gamma_\sigma = 675$ MeV. A broad resonant structure below 900 MeV in the $S$-wave in $\pi^- \pi^+ \to \pi^- \pi^+$ is required to saturate the Adler-Weisberger sum rule for $\pi^-  \pi^+ \to \pi^- \pi^+$ scattering and to produce the equality $m_\sigma = m_\rho$. A narrow resonance like $\sigma(770)$ will not do it. But then, what does saturate the Adler-Weisberger sum rule for $\pi^- \pi^+ \to \pi^- \pi^+$ scattering?\cite{weinberg01}

We suggest that the solutions $(1, \overline 1)$ and $(2, \overline 1)$ provide the answer. In these two solutions the narrow $\sigma(770)$ manifests itself as a broad resonant structure at $\sim$ 720 MeV in the flip helicity amplitude $S_1$. This broad resonant structure will appear also in the $S$-wave in $\pi^- \pi^+ \to \pi^- \pi^+$ and we conjecture that it will saturate the Adler's sum rule as required by Weinberg's mended symmetry.

Weinberg's mended symmetry then selects the solution combinatins $(1, \overline 1)$ and $(2, \overline 1)$. When combined with the Ellis-Lanik relation, the preferred solution is $(1, \overline 1)$. From Table I we see that the corresponding $m_\sigma = 769 \pm 13$ MeV compares well with Particle Data Group value  $m_\rho = 769.3 \pm 0.8$ MeV\cite{PDG00}. In the following we thus refer to the $\sigma$ resonance as $\sigma(770)$.

A question now arises how a chirally invariant and narrow resonance in the nonflip amplitude $S_0$ can also manifest itself as a broad resonant state breaking chiral symmetry in the flip amplitude $S_1$, i.e. in $\pi^- \pi^+ \to \pi^- \pi^+$ scattering. We suggest that the answer may lie in the interaction of the chirally invariant $\sigma(770)$ with the chirally noninvariant QCD vacuum in the  $\pi^- \pi^+ \to \pi^- \pi^+$ scattering.

It is well-known that QED vacuum participates in electromagnetic processes. For instance, the photon propagator is modified by the vacuum polarization\cite{kaku93}. In Ref.~\cite{novikov81} it is shown that the scalar gluonic current interacts strongly with QCD vacuum. Recently Kharzeev and collaborators demonstrated that QCD vacuum participates in hadron interactions at low as well as at high energies with involvement of the QCD trace 
anomaly\cite{kharzeev99,kharzeev00,kharzeev01}. In general, the involvement of the QCD vacuum may depend on the hadronic process and in $\pi^- \pi^+ \to \pi^- \pi^+$ scattering it could result in the metamorphosis of the narrow and chirally invariant $\sigma(770)$ into a broad resonant structure associated with breaking of the chiral symmetry. 

Chiral symmetry allows for two phases of QCD vacuum: asymmetric and degenerate Nambu-Goldstone phase and a symmetric Wigner-Weyl phase\cite{nowak01}. The interaction of the chirally invariant $\sigma(770)$ with QCD vacuum in  $\pi^- \pi^+ \to \pi^- \pi^+$ scattering could also modify the vacuum itself\cite{lee74}. Brown and Rho used the dilaton-gluonium scalar field $\sigma$ and Ellis-Lanik effective lagrangian to describe the change of QCD vacuum in a dense medium\cite{rho91,rho96,rho02}.

Superconvergence sum rules assumed in Weinberg's mended symmetry are based on analyticity of helicity amplitdes.  Analyticity connects hadron dynamics at low and high energies and is thus used also in amplitude analyses of hadronic reactions. QCD vacuum participates in hadron scattering also at low and high enegies\cite{kharzeev01,shifman92}. The reason why analyticity seems to reflect important aspects of hadron dynamics could be that it originates in some way from the structure of QCD vacuum. 
  
Finally we note that in 1968 when Gilman and Harari published their paper\cite{gilman68}, only $\sigma$ and $\rho$ were assumed to contribute to $\pi^-  \pi^+ \to \pi^- \pi^+$ scattering. Today we know of many more resonances the contribution of which should be taken into account in the saturation of the sum rules each of which becomes more complicated. Also, the presence of these resonances enlarges the set of sum rules equations. We suggest that the sum rules equations factorize into a chain of solvable sets with the Gilman-Harari set of 
$\pi, \sigma, \rho$ and $a_1$ forming the first (lowest mass) set. These sets could in fact be the representations (multiplets) of the Weinberg's mended symmetry. If such is the case, then mended symmetry could provide a new organizing principle of the hadron spectra and effective QCD dynamics.

\section{$\sigma(770)-\rho^0(770)$ degeneracy, pion decay constant and 
$SU(2)$ chiral condensate}

The mass and width of the $\sigma$ resonance in the preferred solution combination $(1, \overline 1)$ are from Table I

\begin{equation}
m_\sigma = 769 \pm13\ {\rm MeV}\ ,\ \Gamma_\sigma = 154 \pm 22\ {\rm MeV}
\end{equation}
\noindent
A comparison with Particle Data Group\cite{PDG00} values for mass and width of the $\rho^0$ meson
\begin{equation}
m_\rho = 769.3 \pm 0.8\ {\rm MeV}\ ,\ \Gamma_\rho = 150.2 \pm 0.8\ {\rm MeV}
\end{equation}
\noindent
indicates a strong $\sigma(770) - \rho^0(770)$ degeneracy. This experimentally observed 
$\sigma - \rho$ degeneracy leads to  some interesting consequences. In the following we set $m_\sigma = m_\rho$ and $\Gamma_\sigma = \Gamma_\rho$.

Using the usual $\rho^0 \pi^- \pi^+$ coupling, the $\rho^0 \to \pi^- \pi^+$ decay width is given by\cite{gilman68}

\begin{equation}
\Gamma_\rho = \Gamma(\rho^0 \to \pi^- \pi^+) = {q^3_\rho \over {12 \pi }} {1 \over {F^2_\pi}}
\end{equation}
\noindent
where $F_\pi$ is the pion decay constant and $q_\rho = {1 \over 2} \sqrt{m^2_\rho -4 m^2_\pi}$ is the pion momentum in the rest frame of $\rho^0$. With $\Gamma(\sigma \to \pi^- \pi^+) = \xi \Gamma_\sigma$, the Ellis-Lanik relation (1.5) then relates the gluon condensate and the pion decay constant

\begin{equation}
G_0 = {1 \over \xi} {2 m^5_\sigma \over {(m^2_\rho - 4 m^2_\pi)^{3 \over 2}}} F^2_\pi
\end{equation}
\noindent
The $SU(2)$ chiral condensate $< {\overline q}q >$ is related to the pion decay constant by the Gell-Mann, Oakes and Renner (GMOR) relation\cite{gellmann68}

\begin{equation}
m^2_\pi F^2_\pi = (m_u + m_d)  < {\overline q}q> 
\end{equation}
\noindent
The new measurements of $K^+_{e4}$ decay by E865 Collaboration at BNL\cite{pislak01} have been used recently to verify the validity of the GMOR relation\cite{colangelo00}. Combining (4.4) and (4.5) we get a new relation for chiral condensate in terms of the gluon condensate   

\begin{equation}
(m_u + m_d) <{\overline q} q > = \xi m^2_\pi {{(m^2_\rho - 4 m^2_\pi)^{3 \over 2}} \over {2 m^5_\sigma}} G_0
\end{equation}
\noindent
The new relations (4.4) and  (4.6) are a nontrivial consequence of Ellis-Lanik effective lagrangian which is coding the breaking of scale symmetry and confinement in QCD and the CERN measurements of pion production on polarized target.

With PDG value $F_\pi$ = 92.4 MeV, the relation (4.4) gives the following value for the gluon condensate

\begin{equation}
\xi = {2 \over 3}\ : \ {\rm G_0} = 0.0187\ {\rm GeV}^4
\end{equation}
\noindent
With PDG average value $m_u + m_d$ = 8 MeV and using $G_0$ from (4.7) we obtain
from relation (4.6) a value of chiral condensate

\begin{equation}
\xi = {2 \over 3}\ : \ <{\overline q}q> = 0.0208 \ {\rm GeV}^3 = ( 275 {\rm MeV})^3
\end{equation}
\noindent
The value of $G_0$ is in agreement with the value from Ellis-Lanik relation for the solution combination $(1, \overline 1)$ shown in Table I and in a reasonable agreement with the accepted value (2.1). The value of the QCD scale  $< {\overline q}q > = (\Lambda_{QCD})^3$ from (4.8) is in agreement with recent expectations. 

It is interesting to calculate the gluon and chiral condensates for $\xi = 1$

\begin{equation}
 \xi = {1}\ : \ {\rm G_0} = 0.0125\ {\rm GeV}^4
\end{equation}

\begin{equation}
 \xi = {1}\ : \ <{\overline q}q> = 0.0311 \ {\rm GeV}^3 = ( 315 {\rm MeV})^3
\end{equation}
\noindent
Comparing the values (4.7) and (4.9) for the gluon condensate $G_0$ we notice that the value (4.9) for $\xi$ = 1 is in a better agreement with the accepted value $G_0$ = 0.0117 $\pm$ 0.0051 GeV$^4$ in (2.1). We note that the values of $G_0$ calculated from Ellis-Lanik relation assuming $\xi = 1$ are also closer to the accepted value of $G_0$ in (2.1) but still selecting solutions $(1, \overline 1)$ and $(1, \overline 2)$ with $G_0$ = 0.0116 $\pm$ 0.0028 GeV$^4$ and $G_0$ = 0.0122 $\pm$ 0.0056 GeV$^4$, respectively. It appears that these improved agreements with the accepted value of gluon condensate require an additional factor of ${2 \over 3}$ in the Ellis-Lanik relation (1.5) and in relations (4.4) and (4.6). However we note that $\sigma(770)$ is suppressed in $S$-wave intensity in $\pi^- p \to \pi^0 \pi^0 n$ at 18.3 GeV/c\cite{gunter01}. This suggests that $\Gamma (\sigma \to \pi^0 \pi^0) \to 0$ so that $\xi \to 1$ and improved agrrements follow. An additional term in the Ellis-Lanik effective lagrangian may thus be required by the $\pi^0 \pi^0$ production data to suppress or cancel the $\sigma \pi^0 \pi^0$ coupling.

\section{Summary and Remarks}

CERN measurements of $\pi^- p \to \pi^- \pi^+ n$ on polarized target enable experimental determination of partial wave production spin amplitudes. The measured $S$-wave transversity amplitudes provide evidence for existence of a narrow scalar resonance $\sigma(770)$. The assumption of analyticity of production amplitudes in dipion mass allows the deterimation of the $S$-wave helicity amplitudes $S_0$ and $S_1$.

The unitarity in $\pi^- \pi^+ \to \pi^- \pi^+$ selects "down" solutions $(1, \overline 1)$, $(2, \overline 1)$, $(1, \overline 2)$ and $(2, \overline 2)$. Ellis-Lanik relation selects solutions $(1, \overline 1)$ and $(1, \overline 2)$ while Weinberg's mended symmetry selects solutions $(1, \overline 1)$ and $(2, \overline 1)$. The combined preferred solution $(1, \overline 1)$ has the $\sigma$ meson mass and width degenerate with $\rho^0(770)$. The agreement with Ellis-Lanik relation imparts the $\sigma(770)$ resonance with a dilaton-gluonium interpretation and relates its existence to breaking of scale symmetry in QCD. The QCD vacuum participates in hadron reactions and the interaction of chirally invariant and narrow $\sigma(770)$ with chirally noninvarinat QCD vacuum in $\pi^- \pi^+ \to \pi^- \pi^+$ scattering  is proposed as the mechanism for its metamorphosis into a broad resonant  $S$-wave structure in  $\pi^- \pi^+ \to \pi^- \pi^+$ related to spontaneous breaking of chiral symmetry.  The observed $\sigma  - \rho^0$ degeneracy leads to two new relations for pion decay constant and $SU(2)$ chiral condensate which are both well satisfied experimentally. 

Scalar mesons are probes of QCD vacuum\cite{bijmens00} and the complexity of scalar states reflects the complex structure of the QCD vacuum. There is an emerging evidence for a $\sigma(600)$ resonance from tau decays 
$\tau^- \to a_1^- \nu \to \sigma \pi^- \nu \to \pi^0 \pi^0 \pi^- \nu$\cite{asner99}, from 
$D^+ \to \pi^+\pi^+ \pi^-$ decays\cite{aitala01} and from measurements of 
$\pi^- p \to \pi^0 \pi^0 n$ at 18.3 GeV/c\cite{gunter01}. The CERN measurement of 
$K^+ n \to K^+ \pi^- p$ on polarized target at 5.98 GeV/c suggests existence of a strange scalar 
$\kappa(890)$\cite{svec92b}. Several recent investigations examined the possibility of a scalar nonet below 1 
GeV\cite{beveren86,ishida97,ishida99,black98,black01,oller99,oller99b}. However a 
non-$ {\overline q} q$ structure of $\sigma(600)$ and $\kappa(890)$ has been also 
proposed\cite{steele00,narison02,svec92b}. High statistics measurements of  $\pi^- p \to \pi^0 \pi^0 n$ and $K N \to K \pi N$ on polarized targets\cite{svec97b,svec92b} would help clarify the evidence for these scalar states.

Due to its dilaton-gluonium nature, the $\sigma(770)$ may play a special and a broader role.
Estimate of the cosmological constant from QCD scale anomaly yields a value in a remarkable agreement with the latest astrophysical data\cite{schutzhold02}. A dilatonic nature of  the quintessence has been recently proposed by Damour, Piazza and 
Veneziano\cite{damour02}. Bulk dilaton field in AdS brane world models can be stabilized on the brane and allow for an interaction of the dilaton field with the standard matter through its fluctuation corresponding to an observable massive dilaton\cite{zhuk02}. QCD/string duality stemming from the AdS/CFT correspondence\cite{maldacena98,maldacena00} in AdS brane world model predicts the lowest mass of the scalar gluonium $0^{++}$\cite{boschi02}

\begin{equation}
 m_g(0^{++}) = {1 \over{gN\alpha'}} \sim {\alpha' \over{R^4}}
\end{equation}
\noindent
where $g$ is the string coupling constant, $1/ \sqrt{\alpha'}$ is the string energy scale, $N$ is the number of branes and $R$ is the AdS radius. It is inviting to identify the predicted $m_g(0^{++})$ with the mass of $\sigma(770)$. The resonance parameters of $\sigma(770)$ may then be related to large scale structure and evolution of the Universe. 

Lattice QCD calculations in pure-gauge theory predict large masses for $0^{++}$ gluonium. Recent calculations predict $m_g = 1730 \pm 80$ MeV\cite{morningstar99}. Lower masses are obtained in lattice calculations with scalar glueball - scalar meson mixing in a quenched approximation\cite{mcneile01}. Implementing chiral symmetry in lattice calculations can also modify the predicted mass of the scalar gluonium\cite{hernandez02}. Conformal symmetry may be another additional ingredient in lattice calculations that may bring the predicted gluonium mass closer to the 770 MeV inferred from the CERN measurements of pion production on polarized target.

The $\sigma - \rho^0$ degeneracy leads to observable effects in hadron reactions. The degeneracy implies that the $\sigma(770)$ Regge trajectory $\alpha_\sigma$ is a daugther trajectory of the $\rho(770)$ Regge trajectory $\alpha_\rho$. Such $\sigma$ exchange Regge amplitude is required to describe the observed anomalous energy dependence of polarizations in $pp$ and $np$ elastic scattering and the deviation from the mirror symmetry in polarizations in $\pi^- p$ and $\pi^+ p$ elastic scattering at intermediate 
energies\cite{navelet76a,navelet76b}. This indicates that polarization measurements of hadron reactions even at intermediate energies may provide a testing ground for the renewed efforts to develop Regge amplitudes from AdS/CFT and QCD/string duality\cite{peschanski02}. The Regge behaviour of helicity amplitudes is essential to superconvergence sum rules assumed in Weinberg's mended symmetry\cite{weinberg90}.  The relationship between nonlinearly realized conformal invariance developed by Salam and Strathdee\cite{salam69} and the breaking of chiral symmetry and its connection to mended symmetry was first discussed by Beane and van Kolck in 1994\cite{beane94}.  

A remarkable feature of hadron collisions is the conversion of the kinetic energy of colliding hadrons into the matter of produced particles. The pion production is the simplest of these matter creation processes. The measurements of pion production on polarized targets access the production process on the fundamental level of spin amplitudes rather than spin averaged cross-sections. Our results emphasize the need for a dedicated and systematic study of various production processes on the level of spin amplitudes measured in experiments with polarized targets. Such "amplitude spectroscopy" will be feasible at high intensity hadron 
facilities\cite{svec88,svec93}. The first high intensity hadron facility will be the Japan Hadron Facility at KEK. It will become operational in 2007 and the formation of its experimental program is now in progress\cite{jhf}.

\acknowledgements
I wish to thank V. Elias for his interest and helpful e-mail correspondence concerning QCD sum rules, scale invariance and $\sigma$ meson. I thank M. Rho for bringing to my attention the role of $\sigma$ meson in the change of QCD vacuum in a dense medium and to 
references\cite{rho91,rho96,rho02}. My special thanks go to S. Weinberg for stimulating and encouraging e-mail correspondence concerning $\sigma$ meson and mended symmetry.

\begin{figure}
\caption{The measured transversity amplitudes $|\overline S|^2\Sigma$ and $|S|^2\Sigma$. The curves are the best simultaneous fits to four solution combinations using analyticity based parametrization of transversity amplitudes $\overline S$ and $S$. The Fits A and B yield virtually identical curves.}\label{fig 1}
\end{figure}

\begin{figure}
\caption{Helicity nonflip amplitude $S_0$ and helicity flip amplitude $S_1$ for Fits A and B in the four solution combinations $(1, \overline 1)$, $(2, \overline 1)$, $(1, \overline 2)$ and $(2, \overline 2)$. The fit $A_m$ corresponds to the total absence of $\sigma(770)$ in the helicity flip amplitude $S_1$.}\label{fig 2}
\end{figure}

\begin{table}
\caption{Mass $m_\sigma$ and width $\Gamma_\sigma$ from simultaneous fits to amplitudes $|\overline S|^2\Sigma$ and $|S|^2\Sigma$ in four combinations of solutions of $|\overline S|^2\Sigma$ and $|S|^2\Sigma$. The gluon condensate $G_0$ and the width $\Gamma(\sigma \to \gamma \gamma)$ are calculated from Ellis -Lanik relations (1.5) and (1.6) with $\xi = {2 \over 3}$.}\label{table1}
\begin{tabular}{ccccc}
Fit & $(1, \overline 1)$ & $(2, \overline 1)$ & $(1, \overline 2)$ & $(2, \overline 2)$ \\
$m_\sigma(MeV)$ & 769 $\pm$ 13 & 774 $\pm$ 12 & 787 $\pm$ 19 & 780 $\pm$ 18\\
$\Gamma_\sigma(MeV)$ & 154 $\pm$ 22 & 121 $\pm$ 23 & 165 $\pm$ 45 & 126 $\pm$ 40\\
$G_0(GeV^4)$ & 0.0174 $\pm$ 0.0042 & 0.0227 $\pm$ 0.0063 & 0.0183 $\pm$ 0.0084 & 0.0228 $\pm$ 0.0074\\
$\Gamma_{\gamma \gamma}(eV)$ & 16.5 & 12.9 & 17.6 & 13.5\\
\end{tabular}
\end{table}

\pagebreak
\pagestyle{empty}
\begin{center}
\begin{figure}
\centerline{\epsfysize=7.5in\epsfbox{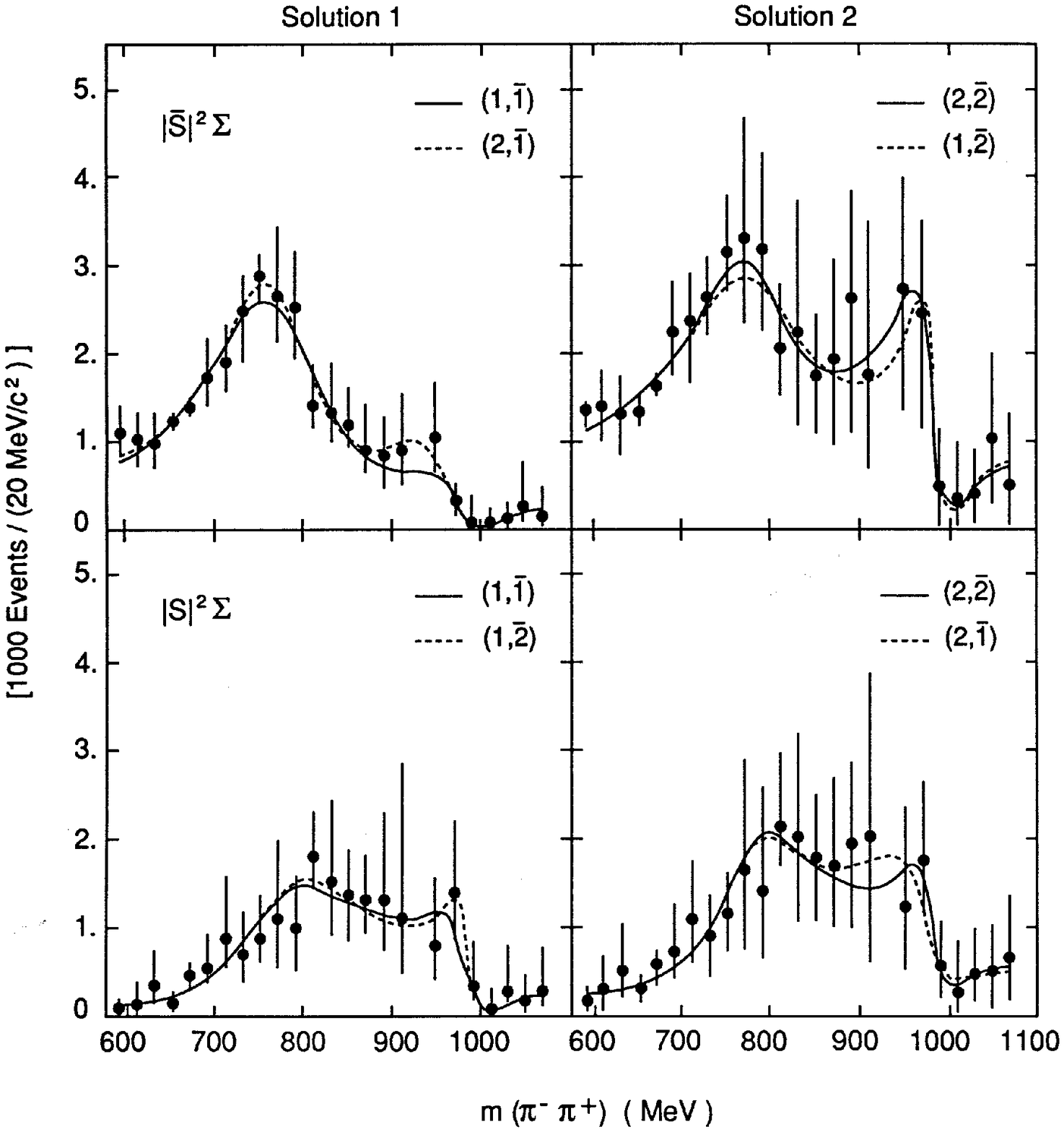}}
\bf Figure 1
\end{figure}
\end{center}
\pagebreak
\begin{center}
\begin{figure}
\centerline{\epsfysize=7.5in\epsfbox{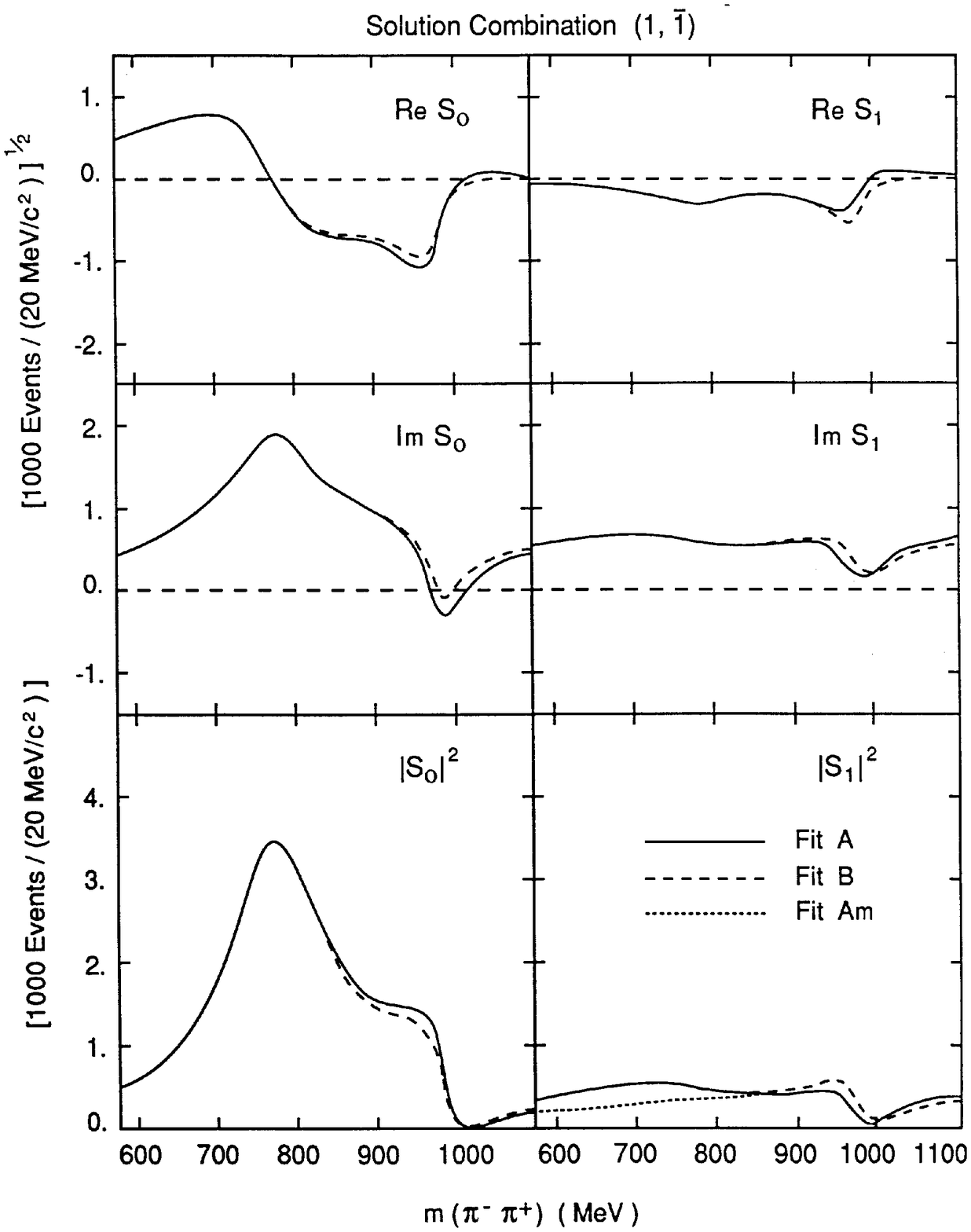}}
\bf Figure 2a
\end{figure}
\end{center}
\pagebreak
\begin{center}
\begin{figure}
\centerline{\epsfysize=7.5in\epsfbox{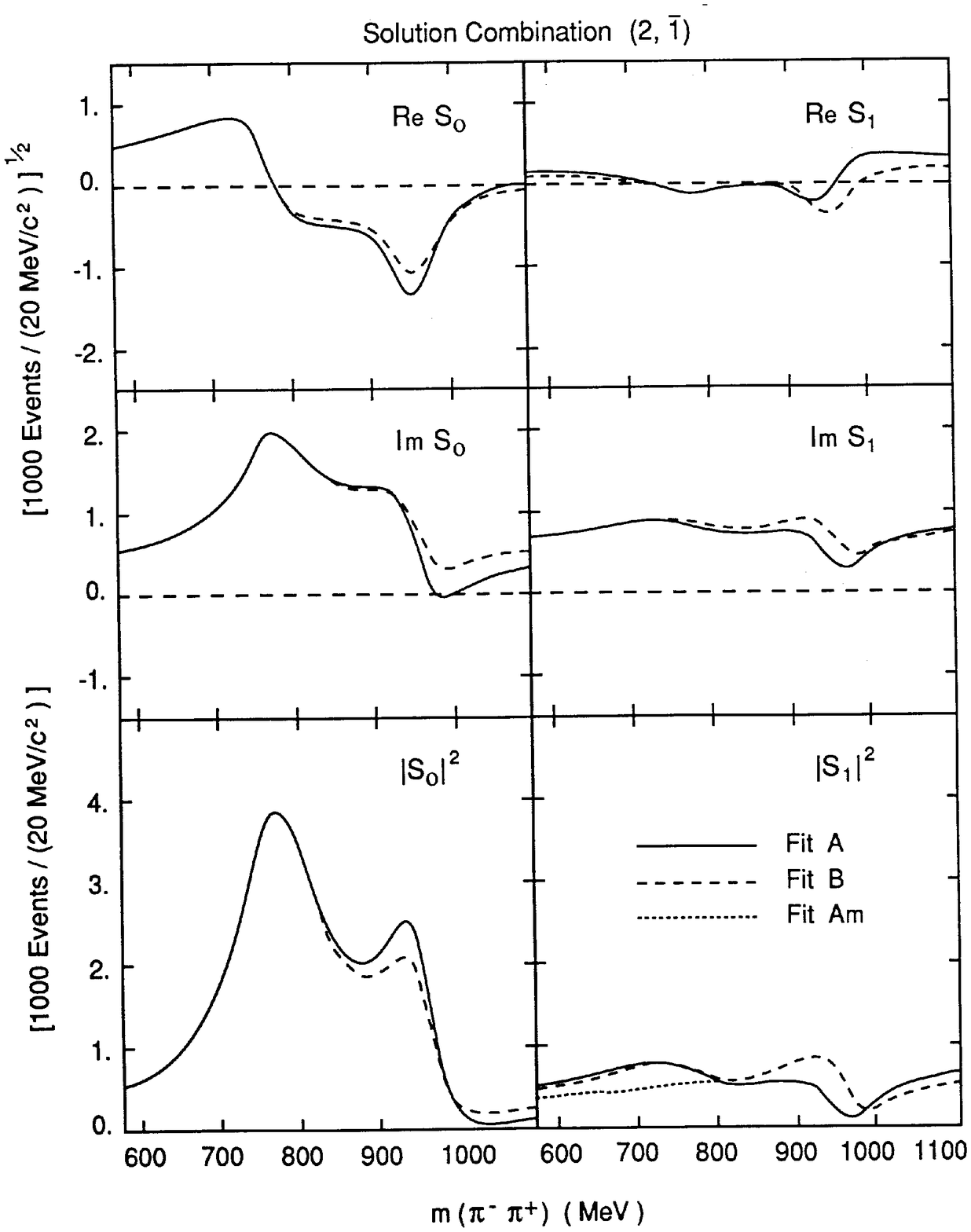}}
\bf Figure 2b
\end{figure}
\end{center}
\pagebreak
\begin{center}
\begin{figure}
\centerline{\epsfysize=7.5in\epsfbox{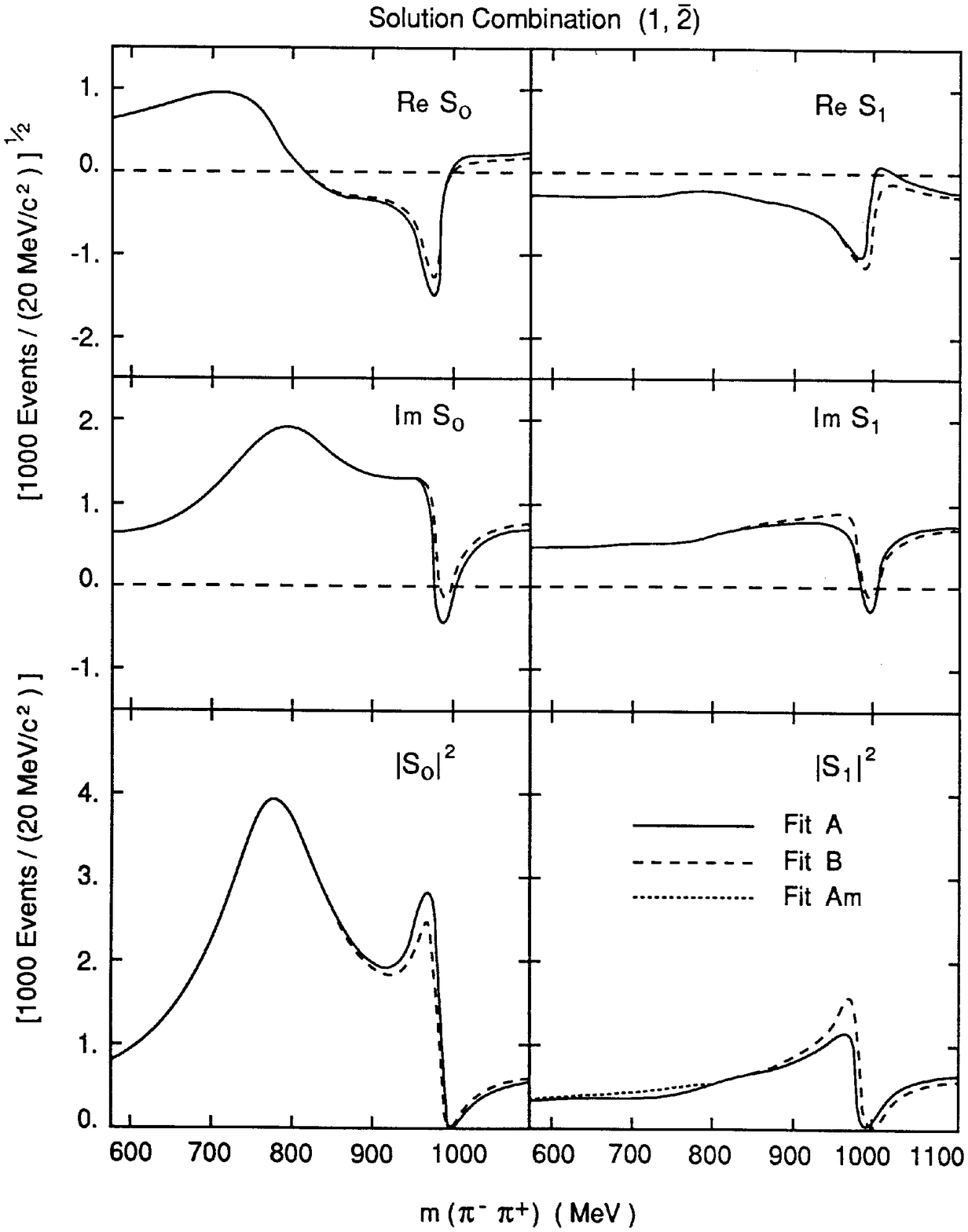}}
\bf Figure 2c
\end{figure}
\end{center}
\pagebreak
\begin{center}
\begin{figure}
\centerline{\epsfysize=7.5in\epsfbox{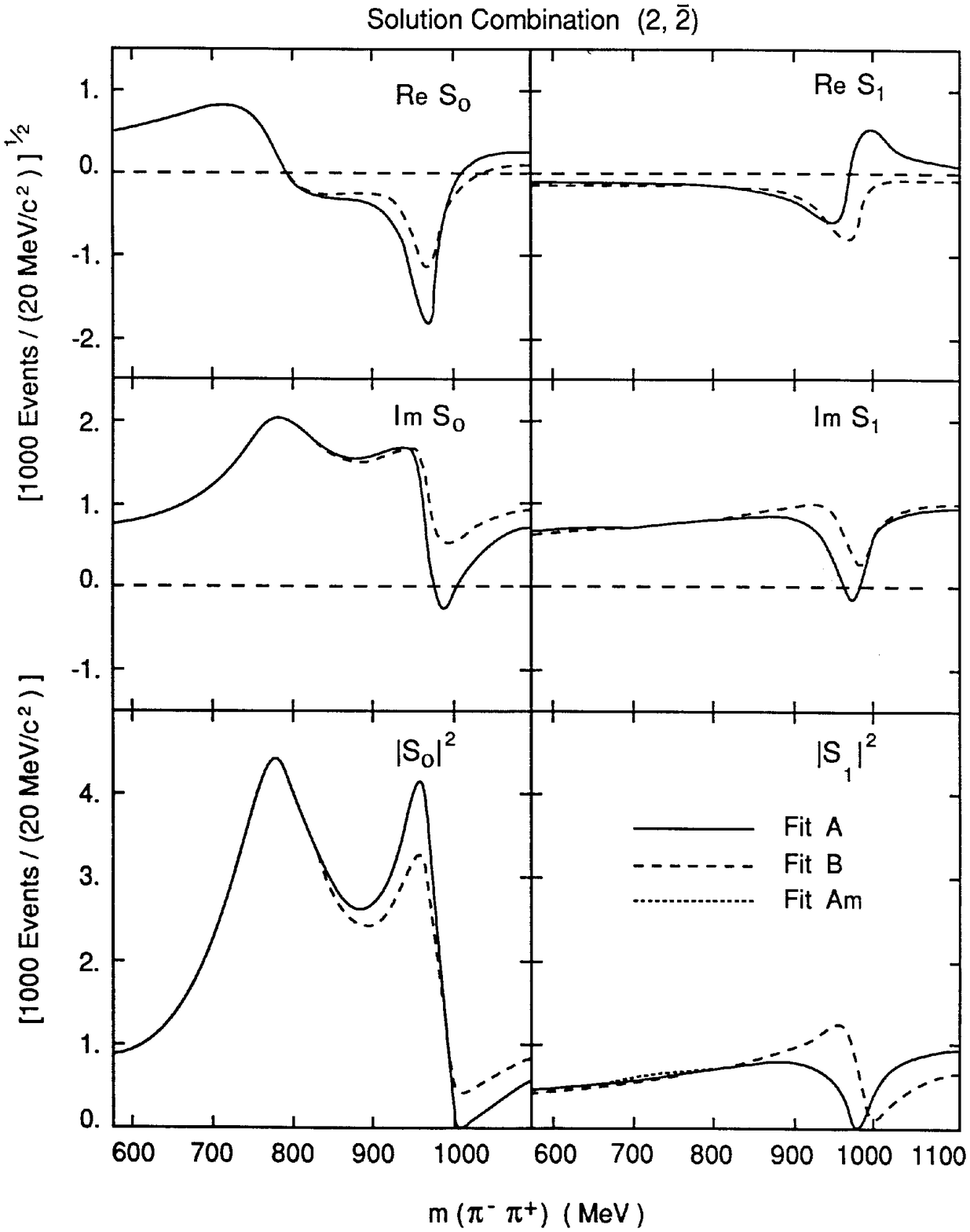}}
\bf Figure 2d
\end{figure}
\end{center}
\end{document}